%% file: article.tex
\documentclass[aps,prb,citeautoscript,twocolumn,longbibliography,superscriptaddress]{revtex4-1}
\usepackage[T1]{fontenc}
\usepackage{latexsym}
\usepackage{graphicx} 
\usepackage{epstopdf}
\usepackage{amsmath}  
\usepackage{amssymb}
\usepackage{comment}
\usepackage{cases}
\usepackage{lipsum}
\usepackage{float}
\usepackage{tikz}
\usepackage{makecell}
\usepackage{bbold}

\usepackage{soul}
\usepackage{color}
\usepackage[breaklinks=true]{hyperref}
\usepackage{xcolor}

\begin{document}

\title{Tuning topological superconductivity within the $t$-$J$-$U$ model of twisted bilayer cuprates}

  \author{Maciej Fidrysiak}
  \email{maciej.fidrysiak@uj.edu.pl}
  \affiliation{Institute of Theoretical Physics, Jagiellonian University, ul. {\L}ojasiewicza 11, 30-348 Krak{\'o}w, Poland }
  \author{Bartłomiej Rzeszotarski}
  \affiliation{Institute of Theoretical Physics, Jagiellonian University, ul. {\L}ojasiewicza 11, 30-348 Krak{\'o}w, Poland }
  \author{J{\'o}zef Spa{\l}ek}%
  \email{jozef.spalek@uj.edu.pl}
  \affiliation{Institute of Theoretical Physics, Jagiellonian University, ul. {\L}ojasiewicza 11, 30-348 Krak{\'o}w, Poland }


\begin{abstract}
  We carry out a theoretical study of unconventional superconductivity in twisted bilayer cuprates (TBC) as a function of electron density and layer twist angle. The bilayer $t$-$J$-$U$ model is employed and analyzed within the framework of a generalized variational wave function approach in the statistically-consistent Gutzwiller formulation. The constructed phase diagram encompasses both gapless $d$-wave state (reflecting the pairing symmetry of untwisted copper-oxides) and gapped $d+\mathrm{e}^{i\varphi}d$ phase that breaks spontaneously time-reversal-symmetry (TRS) and is characterized by nontrivial Chern number. We find that $d+\mathrm{e}^{i\varphi}d$ state occupies a non-convex butterfly-shaped region in the doping vs. twist-angle plane, and demonstrate the presence of previously unreported reentrant TRS-breaking phase on the underdoped side of the phase diagram. This circumstance supports the emergence of topological superconductivity for fine-tuned twist angles in TBC away from $45^\circ$. Our analysis of the microscopically derived Landau free energy functional points toward sensitivity of the superconducting order parameter to small perturbations close to the topological state boundary.
\end{abstract}

\maketitle


\section{Introduction}
\label{section:introduction}

Strong electronic correlations in condensed-matter materials are the driving force of a number of unique phenomena, among them high-temperature superconductivity (SC), pseudogap state, and non-Fermi liquid behavior. Unambiguous microscopic clarification of those and related effects is challenging and remains one of the major endeavors in condensed matter physics (cf. Ref.~\citenum{SpalekPhysRep2022} for review). Successful execution of this task relies on the supply of diversified experimental data for broad range of parameters and energy scales to provide the testing ground for contemporary theoretical models. A new paradigm for an \emph{ad hoc} engineering of highly-tunable strongly correlated electron systems, offering such an insight, appeared following experimental realization of unconventional superconductivity and Mott insulating states in twisted bilayer graphene \cite{CaoNature2018_1,CaoNature2018_2}. In this case, the interlayer twist-angle serves as the parameter providing control over electronic bandwidth, and thus allows to drive the otherwise moderately correlated system into the strongly correlated regime. Analogous tunable strongly-correlated states have been also reported for twisted transition metal dichalcogenides \cite{WangNatMater2020}.

Recent progress in fabrication of high-temperature SC $\mathrm{Bi_2Sr_2CaCu_2O_{8+\mathit{x}}}$ (Bi-2212) thin-films \cite{ZhaoPhysRevLett2019,LiaoNanoLett2018} and monolayers, \cite{YuNature2019} with properties comparable to the bulk crystals, opens up new possibilities to explore physics of strong electronic correlations in two dimensions. Specifically, introducing of a twist between the copper-oxide SC layers has been proposed to stabilize high-temperature $d+\mathrm{e}^{i\varphi}d$ SC with nonzero phase $\varphi$ between layer order parameters, that spontaneously breaks time-reversal-symmetry (TRS) and hosts nontrivial topology \cite{CanNatPhys2021}. This might be regarded as generalization of pure $d+id$ pairing and a concrete realization of the earlier proposals of TRS-breaking SC in Josephson junctions, composed of superconductors with distinct pairing symmetries (e.g., $d_{x^2-y^2}$ and $d_{xy}$) \cite{YangPhysRevB2018}. Subsequent theoretical studies of the Hubbard \cite{LuPhysRevB2022} and $t$-$J$-model \cite{SongPhysRevB2022} on twisted bilayer square lattice support this scenario, yet the resultant SC phase diagrams vary substantially, depending on the microscopic Hamiltonian and approximation employed, encompassing topological SC with large Chern numbers ($C = 2, 4, 8$) or topologically trivial state ($C = 0$). Experimental search for possible formation of TRS-breaking SC in Bi-2212 \cite{ZhuPhysRevX2021,LeeNanoLetters2021,MartiniMaterToday2023,ArXiv:2108.13455} and $\mathrm{Bi_2Sr_{2-\mathit{x}}La_\mathit{x}CuO_{6+\mathit{y}}}$ (Bi-2201) \cite{WangNatCommun2023} provides ambiguous evidence regarding $d+id$ pairing, and points toward prevalence of isotropic pairing component. Systematic characterization of the superconductivity across the phase diagram of twisted bilayer cuprates (TBC) is thus desired.

We carry out analysis of unconventional SC for TBC in broad range of twist angle and hole doping, from underdoped to overdoped regime. A generalized $t$-$J$-$U$ model on twisted square lattice is employed. This Hamiltonian incorporates both on-site Coulomb repulsion and antiferromagnetic exchange on equal footing, and has been previously utilized for a quantitative description of equilibrium and dynamic properties of untwisted materials \cite{SpalekPhysRep2022}. The analysis is carried out within the framework of \textbf{S}tatistically-consistent \textbf{G}utzwiller \textbf{A}pproximation (SGA), constituting a finite-temperature extension of the variational wave function scheme that is applicable to large supercells, emerging in moir\'{e} systems. The obtained doping vs. twist-angle phase diagram encompasses both plain gapless $d$-wave and topological TRS-breaking $d+\mathrm{e}^{i\varphi}d$ SC. Those states are separated by a series of quantum phase transitions concealed within the high-$T_c$ SC dome inherited from the untwisted cuprates.

The $d+\mathrm{e}^{i\varphi}d$ state is stabilized in a narrow layer twist angle regime around $\theta = 45^\circ$, in agreement with former theoretical studies \cite{CanNatPhys2021,LuPhysRevB2022}. However, we reveal previously unreported reentrant $d+\mathrm{e}^{i\varphi}d$ superconductivity emerging as a function of layer twist angle in underdoped systems. Multiple characteristics of SC state are analyzed throughout the high-$T_c$ phase diagram, including SC order parameter, equilibrium value of SC relative phase factor between the layers, and energy gap in the quasiparticle spectrum. In particular, we demonstrate that the onset of $d+\mathrm{e}^{i\varphi}d$ SC is assisted by opening of the energy gap, yet those two quantities follow distinct doping dependence; for certain values of layer twist angle, we identify a double-dome shaped gap as a function of electronic density (band filling). Finally, we extend our analysis of equilibrium SC properties and study microscopically-derived Landau free energy functional, which encodes information about the stiffness of paired state. Close to the onset of TRS-breaking SC, free energy becomes extremely flat as a function of the relative SC phase factor between layers. This circumstance points toward sensitivity of the $d+\mathrm{e}^{i\varphi}d$ order parameter to small perturbations, such as chemical disorder and interface imperfections, and might rationalize ambiguous structure of the SC order parameter reported experimentally for twist angles close to $\sim 45^\circ$.

The paper is organized as follows. In Sec.~\ref{section:model} we introduce the bilayer $t$-$J$-$U$ model and in Sec.~\ref{section:order_parameter} we discuss the symmetry of the SC order parameter and define relevant quantities. In Sec.~\ref{section:results} we present the results, encompassing SGA doping vs. twist angle phase diagram, as well as the analysis of microscopically derived Landau free-energy landscape vs. $d+\mathrm{e}^{i\varphi}d$ pairing amplitude. Summary and discussion is given in Sec.~\ref{section:summary}. Appendices~\ref{appendix:SGA}-\ref{appendix:landau_free_energy} provide relevant details of our approach and multiple consistency checks of employed methodology.

\section{Model and method}
\label{section:model}

\begin{figure*}
  \centering
  \includegraphics[width=1\linewidth]{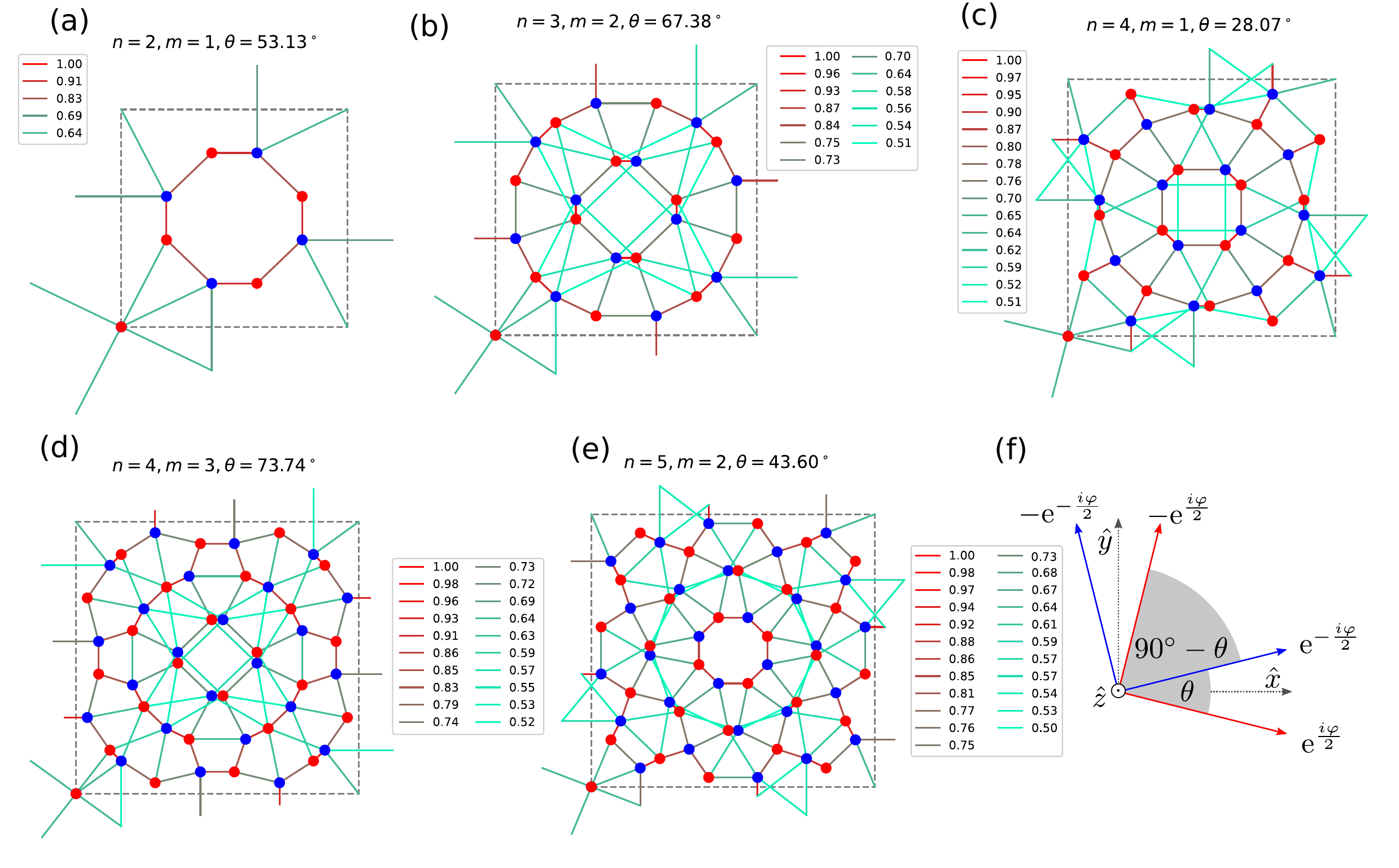}
  \caption{Top view of the unit cell for twisted-bilayer square lattice. Blue- and red circles are copper sites in the lower- and upper-layers, respectively, and links represent interlayer  hybridization, $V_{ij}$, here given in the units of $V_0$ [cf. Eq.~\eqref{eq:tunneling}]. Only links corresponding to hybridization $V_{ij} \geq 0.5 V_0$ are marked for clarity (magnitudes of $V_{ij}/V_0$ are detailed in the legend). Note that the copper sites in neighboring layers in the left-bottom corner of the unit cell are positioned exactly on top of each other, and blue circle is not visible. Panels (a)-(e) correspond to twist angles $\theta = 53.13^\circ$, $67.38^\circ$, $28.07^\circ$, $73.74^\circ$, and $43.60^\circ$, respectively. Panel (f) illustrates the symmetry of $d+\mathrm{e}^{i\varphi}d$ superconducting order parameter in twisted square lattice. Blue- and red coordinate axes correspond to the layers, twisted by and angle $\pm\theta/2$ with respect to the original coordinate frame (dotted axes). Phases of the layer order parameters $\pm \exp(\pm i \varphi/2)$ along the directions parallel to the axes are also marked, as explained in the text. As follows from (f), system at twist angle $\theta$ may be transformed into that twisted by complementary $\theta^\prime = 90^\circ - \theta$ by means of a $45^\circ$ lattice rotation along the out-of-plane $\hat{z}$ direction, combined with the interchange of $A$ and $B$ planes. The data for complementary angles $\theta^\prime = 36.87^\circ$, $61.93^\circ$, $16.26^\circ$, and $46.40^\circ$ are thus obtained by transforming those for supercells (a)-(e), and there is no need to consider them explicitly.}
  \label{fig:lattice}
\end{figure*}

For certain (commensurate) twist angles, the twisted bilayer square lattice forms periodic superstructure. Those configurations are characterized by two integers, $n$ and $m$, such that the layer twist angle reads $\theta = 2 \cdot \arctan(m/n)$. The number of copper atoms per unit cell reads then $N_\mathrm{cell} = 2\cdot (n^2 + m^2)$, with the factor of two reflecting the number of layers. The resultant superstructure forms a square with lattice spacing $a = \sqrt{n^2 + m^2} a_0$, where $a_0$ denotes the in-plane Cu-Cu distance. Hereafter, we set the total number of copper sites to $N_\mathrm{Cu} \approx 80\,000$ which, as we verified, is sufficient to reliably represent thermodynamic-limit situation. Note that it is not possible to form periodic lattices with exactly $N_\mathrm{Cu} \equiv 80\,000$ sites for all considered twist angles, as the unit-cell dimensions are irrational numbers in general. In Table~\ref{tab:bilayer_superstructures} we list the values of $n$ and $m$ addressed in the present study, as well as lattice sizes and other relevant parameters. The unit cells of resulting superstructures are illustrated in Figure~\ref{fig:lattice}(a)-(e), with the exception of the trivial case without a twist ($\theta = 0^\circ$) and those marked as \emph{not included} in Table~\ref{tab:bilayer_superstructures}. Blue- and red circles represent Cu sites in layers $A$ and $B$, respectively, and links visualize hybridization between layers. Copper sites in the left-bottom corner of the unit cell are positioned exactly on top of each other so that the blue circle is not visible.

\begin{table}
  \centering
  \caption{Characteristics of the twisted square-lattice bilayer for selected twist angles, $\theta$. Commensurate configurations are determined by two integers, $n$ and $m$, such that $\theta = 2 \cdot \arctan(m/n)$. $N_\mathrm{cell}$ is the number of atoms in the unit cell, and $N_\mathrm{Cu}$ denotes the total number of Cu sites in the lattice (periodic boundary conditions are imposed and lattice dimensions have been selected so that $N_\mathrm{Cu}$ is close to $80\,000$). Superstructure lattice constant is also given, in the units of the Cu-Cu distance $a_0$. Inequivalent lattices used in the present paper are marked in the last column as \emph{included}. Those \emph{not included} are related by symmetry to other superlattices with smaller unit cells, as discussed in the text.}
  \label{tab:bilayer_superstructures}
  \begin{tabular}{cccccccc}
    \hline \\
    $n$& $m$ & $N_\mathrm{cell}$ & $N_\mathrm{Cu}$ & $a/a_0$ & $\theta \,({}^\circ)$ & $90^\circ - \theta \,({}^\circ)$ & included\\
    \hline \hline \\
    1 & 0 & 2 & 80\,000 & 1.000 & 0.00 & 90.00 & yes \\
    2 & 1 & 10 & 79\,210 & 2.236 & 53.13 & 36.87 & yes \\
    3 & 1 & 20 & 79\,380 & 3.162 & 36.87 & 53.13 & no \\
    3 & 2 & 26 & 78\,650 & 3.606 & 67.38 & 22.62 & yes \\
    4 & 1 & 34 & 78\,336 & 4.123 & 28.07 & 61.93 & yes\\
    4 & 3 & 50 & 80\,000 & 5.000 & 73.74 & 16.26 & yes\\
    5 & 1 & 52 & 79\,092 & 5.099 & 22.62 & 67.38 & no \\
    5 & 2 & 58 & 79\,402 & 5.385 & 43.60 & 46.40 & yes \\
    \hline
  \end{tabular}
\end{table}

Remarkably, not all pairs of indices $(n, m)$ result in new superstructures. Layers at twist angle $\theta$ may be transformed into those twisted by $\theta^\prime = 90^\circ - \theta$  by appropriate spatial transformations and are thus physically equivalent in the thermodynamic limit. Small finite-size and boundary condition effects arising in finite geometries, are discussed below. In effect, there is no need to explicitly consider lattices characterized by $n=3, m=1$ and $n=5, m=1$ as those are equivalent to $n=2, m=1$ and $n=3, m=2$ with smaller supercells (those are marked as \emph{not included} in Table~\ref{tab:bilayer_superstructures}; we use them only for test and benchmark purposes). The geometric construction illustrating this equivalence is displayed in Fig.~\ref{fig:lattice}(f). Blue- and red- coordinate axes correspond to the square lattices in layers $A$ and $B$, respectively. The layers are rotated with respect to the original coordinate system $\hat{x}$-$\hat{y}$ (dotted arrows) by $+\theta/2$ (counter-clockwise) and $-\theta/2$ (clockwise), so that the relative angle between layers reads $\theta$. By carrying out a $45^\circ$ counter-clockwise rotation around the out-of-plane $\hat{z}$ axis, followed by layer interchange, we effectively exchange the angles $\theta$ and $90^\circ - \theta$ marked in panel (f). In effect, with a limited number of superstructures, we are able to densely cover the range of twist angles $\theta \in [0, 90^\circ]$ and explore the structure of doping-$\theta$ phase diagrams for correlated lattice Hamiltonians on twisted square lattice.

We employ the $t$-$J$-$U$ model \cite{SpalekPhysRep2022,SpalekPhysRevB2017}, reformulated for the TBC system. It may be regarded as a formal generalization of Hubbard and $t$-$J$ models, and encompasses both of them as particular cases. The Hubbard model is not well suited for SGA and related (e.g., slave-boson) approximation schemes due to lack of explicit exchange interactions. The magnetic exchange processes emerge in subleading orders of respective diagrammatic expansions,\cite{FidrysiakJPCM2018} making the problem intractable for large supercells addressed. In effect, low-order approximation solutions obtained within those three Hamiltonians may differ. The $t$-$J$ and $t$-$J$-$U$ models provide thus a methodological advantage over the complementary Hubbard-model description, allowing to discuss the high-$T_c$ SC already at the saddle-point-solution level. On the other hand, systematic variational wave function studies of both $t$-$J$ and $t$-$J$-$U$ model demonstrate that the $t$-$J$-$U$ model yields better global description of high-$T_c$ cuprates across their phase diagram than the $t$-$J$ model \cite{SpalekPhysRep2022}. In particular, the $t$-$J$-$U$ model reproduces quantitatively experimental doping dependence of the effective masses and Fermi velocity \cite{SpalekPhysRep2022}, as well as correctly accounts for certain subtle SC properties, including a crossover between BCS-like to non-BCS-like regimes. The latter effect occurs at the single meV scale, and is characterized by the sign change of kinetic energy gain at the SC transition \cite{DeutscherPhysRevB2005,LevalloisPhysRevX2016,SpalekPhysRevB2017}. Since the free energy landscape for TBC involves even smaller energy scales, we consider $t$-$J$-$U$ model as an appropriate departure point for addressing unconventional SC in those systems.

An attractive methodological feature of the $t$-$J$-$U$ Hamiltonian is that calculations in restricted (projected) Hilbert space, with doubly-occupied sites excluded, are not necessary. This is because the large value of onsite-Coulomb-repulsion to hopping ratio $\sim 20$ reduces essentially the value of doubly-occupancy probability, $d^2$. Typically $d^2 \lesssim 10^{-2}$, which is much smaller than the weak-coupling-limit value $d^2 \sim 0.2$. Hence, the considered model reflects the principal features of $t$-$J$ model physics. Moreover, the presence of the exchange term $\propto J$ accounts also for additional superexchange channels beyond the Anderson kinetic exchange, which is of too small magnitude for the realistic values of hopping and Coulomb interaction taken for the cuprates. Nevertheless, even in the present strong-correlation regime ($U/|t| \sim 20$) and nonzero doping, the double occupancy probability is small but nonzero. The limit of $t$-$J$ model is thus approached, but not formally reached \cite{ChaoJPhysC1977}. This and other aspects of the approach have been elaborated at length in our previous papers \cite{SpalekPhysRep2022,SpalekPhysRep2022}.

For each of the two layers, $L = A, B$, the respective $t$-$J$-$U$ Hamiltonians read

\begin{align}
  \label{eq:tju-model}
  \hat{\mathcal{H}}_\text{$t$-$J$-$U$}^{(L)} = & \sum_{\substack{\langle i, j\rangle \\ i, j \in L}} \left( t_{ij} \hat{a}^{\dagger}_{i\sigma} \hat{a}_{j\sigma} + \mathrm{H.c.}\right) + \nonumber \\ & \sum_{\substack{\langle i, j\rangle \\ i, j \in L}} J_{ij} \hat{\mathbf{S}}_i \hat{\mathbf{S}}_j  + U \sum_{i \in L}  \hat{n}_{i\downarrow} \hat{n}_{i\uparrow},
\end{align}

\noindent
where $\hat{a}_{i\sigma}$ ($\hat{a}^\dagger_{i\sigma}$) annihilate (create) electrons on lattice site $i$ (note that the operators are not projected as in the $t$-$J$ model \cite{ChaoJPhysC1977}). In Eq.~(\ref{eq:tju-model}) $\langle i, j\rangle$ means that each pair of indices appears only once in the summation, whereas $t_{ij}$ and  $J_{ij}$ denote hopping and exchange integrals, respectively. The Hubbard- and $t$-$J$-model limits are retrieved for $J= 0, U \neq 0$ and $J \neq 0, U \rightarrow \infty$, as discussed above. Hereafter, we retain only hopping integrals between nearest- and next-nearest neighbors, $t = -0.35\,\mathrm{eV}$ and $t^\prime = 0.3 |t|$, and the on-site Coulomb repulsion is set to $U = 20 |t|$. This means that the ratio $U/W$ is large, but not yet in the canonical $t$-$J$-model limit. We consider two values of the antiferromagnetic nearest-neighbor exchange, $J = 0.3 |t|$ and $0.4 |t|$, to assess how the intralayer pairing affects the interlayer SC correlations. Moreover, in our calculations, we retain small finite temperature $T = 10^{-6} |t| / k_B$, where $k_B$ denotes Boltzmann constant. Microscopic parameters in this range have been previously employed to study high-$T_c$ SC \cite{SpalekPhysRevB2017} and interlayer tunneling effects \cite{ZegrodnikPhysRevB2017} in untwisted ($\theta = 0^\circ$) systems.

The layers, $A$ and $B$, are subsequently twisted by an angle $\theta \in [0^\circ, 90^\circ]$, as illustrated in Fig.~\ref{fig:lattice}, and coupled by interlayer tunneling $V_{ij}$ so that the total system Hamiltonian takes the form

\begin{align}
  \label{eq:tju-model-layers}
  \hat{\mathcal{H}} = & \hat{\mathcal{H}}_\text{$t$-$J$-$U$}^{(A)} + \hat{\mathcal{H}}_\text{$t$-$J$-$U$}^{(B)} + \sum_{\substack{i \in B \\ j \in A}, \sigma} \left( V_{ij} \hat{a}^{\dagger}_{i\sigma} \hat{a}_{j\sigma} + \text{H.c.} \right).
\end{align}

\noindent
Whereas $V_{ij}$ generally contains both isotropic- and anisotropic components \cite{MarkiewiczPhysRevB2005,SongPhysRevB2022}, here we restrict to a model situation with pure isotropic and exponentially decaying tunneling

\begin{align}
  \label{eq:tunneling}
  V_{ij} = V_0 \mathrm{e}^{-\frac{ ||\mathbf{r}_i-\mathbf{r}_j|| - d}{r_0}},
\end{align}

\noindent
where $\mathbf{r}_i$ denotes the position of lattice-site $i$, $d$ is interlayer distance, and $r_0$ represents the characteristic decay length of the interaction between layers. Eq.~\eqref{eq:tunneling} is written so that $V_{ij} \equiv V_0$ whenever site $i$ is positioned directly on top of site $j$. In effect, $V_0$ becomes a direct measure of the overall interlayer hopping magnitude. Hereafter, for all the calculations we use $V_0 = -0.2 |t|$, $d = 2.1 a_0$, $r_0 = 0.5 a_0$, with $a_0$ being interlayer Cu-Cu distance. Moreover, to make the problem tractable, we apply a cutoff to interlayer hybridization by discarding all $V_{ij}$ such that $||\mathbf{r}_i-\mathbf{r}_j|| > 3.1 a_0$. The model parameters, employed int he present study, are summarized in Table~\ref{tab:model_parameters}.

\begin{table}
  \centering
  \caption{Summary of the bilayer $t$-$J$-$U$-model parameters, employed in the present study. Except for the value on nearest-neighbor hopping, t, the energies are expressed in the units o |t|. The interlayer distance and tunneling decay length (cf. Eq.~\eqref{eq:tunneling}), are expressed in terms on in-plane Cu-Cu distance, $a_0$. The analysis has been carried out for two values of $J/|t| = 0.3$ and $0.4$.}
  \label{tab:model_parameters}
  \begin{tabular}{ccl}
    \hline \\
    Parameter & Value & Description\\
    \hline \hline \\
    $t$ & $-0.35\,\mathrm{eV}$ & nearest-neighbor hopping\\
    $t^\prime/|t|$ & $0.3$ & next-nearest-neighbor hopping \\
    $U/|t|$ & $20$ & on-site Coulomb repulsion\\
    $J/|t|$ & $0.3$, $0.4$ & exchange interaction \\
    $V_0 / |t|$ & $-0.2$ & interlayer tunneling magnitude\\
    $d/a_0$ & $2.1$ & interlayer distance\\
    $r_0/a_0$ & $0.5$ & tunneling decay length\\
    $k_B T / |t|$ & $10^{-6}$ & temperature \\
    \hline
  \end{tabular}
\end{table}

The model is analyzed within the framework of variational wave function approach in the SGA formulation. In its plain form, varietal method is based on optimization of the energy functional

\begin{align}
  \label{eq:var_en}
E_\mathrm{var} \equiv  \frac{\langle\Psi_\mathrm{var}| \hat{\mathcal{H}} |\Psi_\mathrm{var}\rangle }{ \langle\Psi_\mathrm{var}|\Psi_\mathrm{var}\rangle},
\end{align}

\noindent
where $|\Psi_\mathrm{var}\rangle$ is a variational state accounting for strong electronic correlations on copper sites. SGA provides an extension of the latter to finite temperature, and introduces simplifications that makes the study large lattices and supercells feasible. The methodological details are presented in Appendix~\ref{appendix:SGA}.

\section{Superconductivity and order parameter}
\label{section:order_parameter}

Untwisted high-$T_c$ copper-oxides host $d_{x^2 - y^2}$ superconductivity that is compatible with square-lattice symmetry, and allows the electrons forming Cooper pairs to effectively avoid strong on-site Coulomb interactions. The constituents comprising natural building blocks for the SC order parameter of twisted bilayer are thus two intralayer $d_{x^2-y^2}$-wave order parameters $\Delta_d^{(A)} \equiv \langle\hat{\Delta}_d^{(A)}\rangle$ and $\Delta_d^{(B)} \equiv \langle\hat{\Delta}_d^{(B)}\rangle$, with their respective pairing operators

\begin{align}
  \hat{\Delta}_d^{(A)} = \frac{1}{4 N_\mathrm{Cu}} \sum_{i \in A} \Big (& \hat{a}_{i + \hat{x}_{A} \downarrow} \hat{a}_{i \uparrow} - \hat{a}_{i + \hat{x}_{A} \uparrow} \hat{a}_{i \downarrow} - \nonumber\\ &\hat{a}_{i  \hat{y}_{A} \downarrow} \hat{a}_{i \uparrow} + \hat{a}_{i + \hat{y}_{A} \uparrow} \hat{a}_{i \downarrow} \Big ), \label{eq:delta_a}          \end{align}

\noindent
and

\begin{align}                                       
  \hat{\Delta}_d^{(B)} =  \frac{1}{4 N_\mathrm{Cu}} \sum_{i \in B} \Big ( & \hat{a}_{i + \hat{x}_{B} \downarrow} \hat{a}_{i \uparrow} - \hat{a}_{i + \hat{x}_{B} \uparrow} \hat{a}_{i \downarrow} - \nonumber\\& \hat{a}_{i + \hat{y}_{B} \downarrow} \hat{a}_{i \uparrow} + \hat{a}_{i + \hat{y}_{B} \uparrow} \hat{a}_{i \downarrow} \Big ). \label{eq:delta_b}
\end{align}

\noindent
In Eqs.~\eqref{eq:delta_a}-\eqref{eq:delta_b}, $\hat{x}_A = (\cos \frac{\theta}{2}, \sin \frac{\theta}{2}, 0)$, $\hat{y}_A = (-\sin \frac{\theta}{2}, \cos \frac{\theta}{2}, 0)$, $\hat{x}_B = (\cos \frac{\theta}{2}, -\sin \frac{\theta}{2}, 0)$, $\hat{y}_B = (\sin \frac{\theta}{2}, \cos \frac{\theta}{2}, 0)$ are the basis vectors for sublattices $A$ and $B$, here expressed in the units of the Cu-Cu in-plane distance. The above expressions are normalized by $N_\mathrm{Cu}$ and an additional factor $4$ to take into account four expectation values appearing in the summation. We have found that, for twisted lattice structures displayed in Fig.~\ref{fig:lattice}(a)-(e), the initially imposed $d$-wave intralayer symmetry is preserved throughout SGA self-consistent procedure; no admixture of extended $s$-wave components

\begin{align}
  \hat{\Delta}_s^{(A)} = \frac{1}{4 N_\mathrm{Cu}} \sum_{i \in A} \Big (& \hat{a}_{i + \hat{x}_{A} \downarrow} \hat{a}_{i \uparrow} + \hat{a}_{i + \hat{x}_{A} \uparrow} \hat{a}_{i \downarrow} + \nonumber\\ &\hat{a}_{i  \hat{y}_{A} \downarrow} \hat{a}_{i \uparrow} + \hat{a}_{i + \hat{y}_{A} \uparrow} \hat{a}_{i \downarrow} \Big ), \label{eq:delta_a_swave} \\
  \hat{\Delta}_s^{(B)} =  \frac{1}{4 N_\mathrm{Cu}} \sum_{i \in B} \Big ( & \hat{a}_{i + \hat{x}_{B} \downarrow} \hat{a}_{i \uparrow} + \hat{a}_{i + \hat{x}_{B} \uparrow} \hat{a}_{i \downarrow} + \nonumber\\& \hat{a}_{i + \hat{y}_{B} \downarrow} \hat{a}_{i \uparrow} + \hat{a}_{i + \hat{y}_{B} \uparrow} \hat{a}_{i \downarrow} \Big ) \label{eq:delta_b_swave}
\end{align}

\noindent
is observed, even for largest considered supercells.

The layer order parameters $\Delta_d^{(A)}$ and $\Delta_d^{(B)}$ are generally complex numbers. Since the Hamiltonians for the two layers are equivalent [cf. Eq.~\eqref{eq:tju-model}] and the hybridization $V_{ij}$ acts symmetrically on $A$ and $B$ layers, the order parameter amplitudes may be considered equal $|\Delta_d^{(A)}| = |\Delta_d^{(B)}|$, which we have also verified numerically. In turn, the single parameter $\Delta_\mathrm{SC} \equiv |\Delta_d^{(A)}| = |\Delta_d^{(B)}|$ is hereafter used as a dimensionless measure of superconducting correlations. Moreover, the self-consistently obtained relative phase $\varphi \equiv \arg \left(\Delta_d^{(B)} / \Delta_d^{(A)} \right)$ turns out generally non-trivial, constituting a physically sound quantity. The equilibrium value of $\varphi$ different from $0^\circ$ and $180^\circ$ (modulo $360^\circ$) implies spontaneous breakdown of TRS, and admits nontrivial topology of the SC state. This is caused by the circumstance that under time-reversal operation, the phase is transformed as $\varphi \rightarrow - \varphi$, leading to ground-state degeneracy. With the help of global $U(1)$ gauge transformation, layer order parameters may be generically brought to the form $\Delta_d^{(A)} = |\Delta_d^{(A)}| \exp(-i\varphi/2)$ and $\Delta_d^{(B)} = |\Delta_d^{(B)}| \exp(i\varphi/2)$. The directional dependence of the  order-parameter phase is marked in Fig.~\ref{fig:lattice}(f) next to the respective coordinate axes. Within each of the layers, the phase factor changes sign following spatial rotation by $90^\circ$. The relative phase between $A$- and $B$-layer order parameters is fixed at $\varphi$, which correspond to the $d + \mathrm{e}^{i\varphi}d$ SC.

The above considerations let us relate the order parameters between the superstructures for complementary twist angles ($\theta$ and $\theta^\prime \equiv  90^\circ - \theta$), cf. Table~\ref{tab:bilayer_superstructures}. For the system at the twist angle $\theta$ whose free energy has degenerate minima at $\pm \varphi$, one can carry out the transformation to the $\theta^\prime$ case following the procedure described in Sec.~\ref{section:model}. As follows from Fig.~\ref{fig:lattice}(f), the phase angles $\pm \varphi$ are then transformed into $180^\circ \pm \varphi$. This circumstance substantially reduces the computational cost of evaluating phase diagrams and is utilized below.

\begin{figure}
  \centering
  \includegraphics[width=1\linewidth]{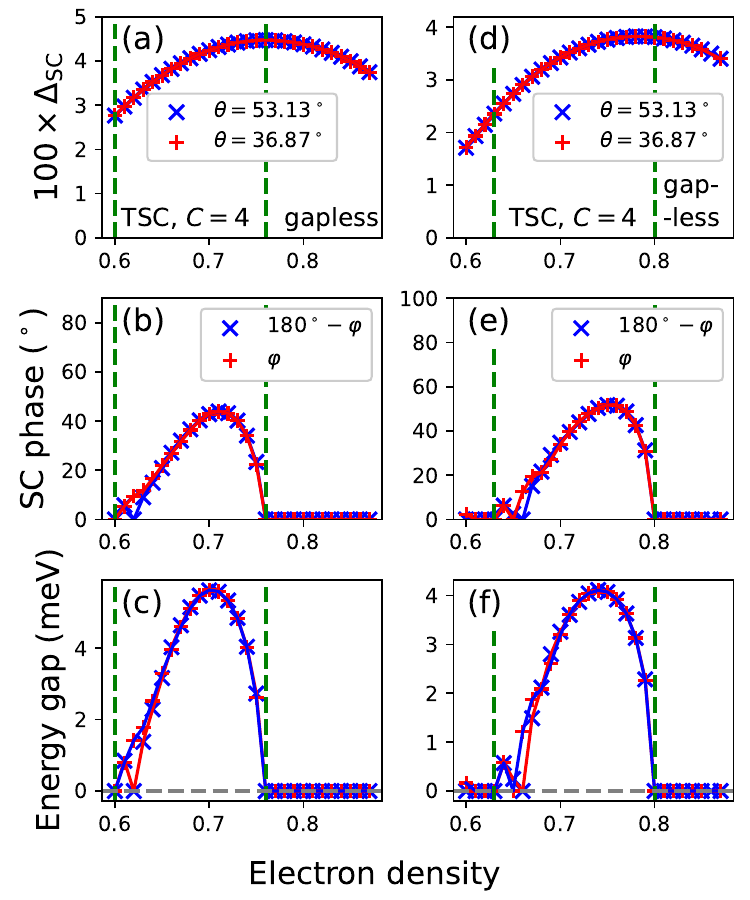}
  \caption{Electron-density (doping) dependence of the calculated order parameter equilibrium value [(a), (d)], relative SC phase between layers, $\varphi$ [(b), (e)], and energy gap in the quasiparticle spectrum [(c), (f)] for two complementary twist angles $\theta = 53.13^\circ$ (blue symbols) and $\theta = 36.87^\circ = 90^\circ - 53.13^\circ$ (red symbols). Those two selections of $\theta$ correspond to different unit cells ($n=2, m=1$ and $n=3, m=1$, respectively, cf. Table~\ref{tab:bilayer_superstructures}), but are related by symmetry. Quantitative agreement of the two-independent simulation results validates our supercell Hamiltonians and approach. Panels (a)-(c) and (d)-(f) correspond to antiferromagnetic exchange $J = 0.4 |t|$ and $J = 0.3 |t|$, respectively. The remaining model parameters are detailed in the text. Vertical green dashed lines mark the quantum phase transitions between gapless $d$-wave and gapped topological SC (TSC) states. The TSC state is characterized by nontrivial Chern number $C=4$.}
  \label{fig:symmetry_test}
\end{figure}

\section{Results}
\label{section:results}

\subsection{Identification of gapped and gapless states}

\begin{figure*}
  \centering
  \includegraphics[width=1\linewidth]{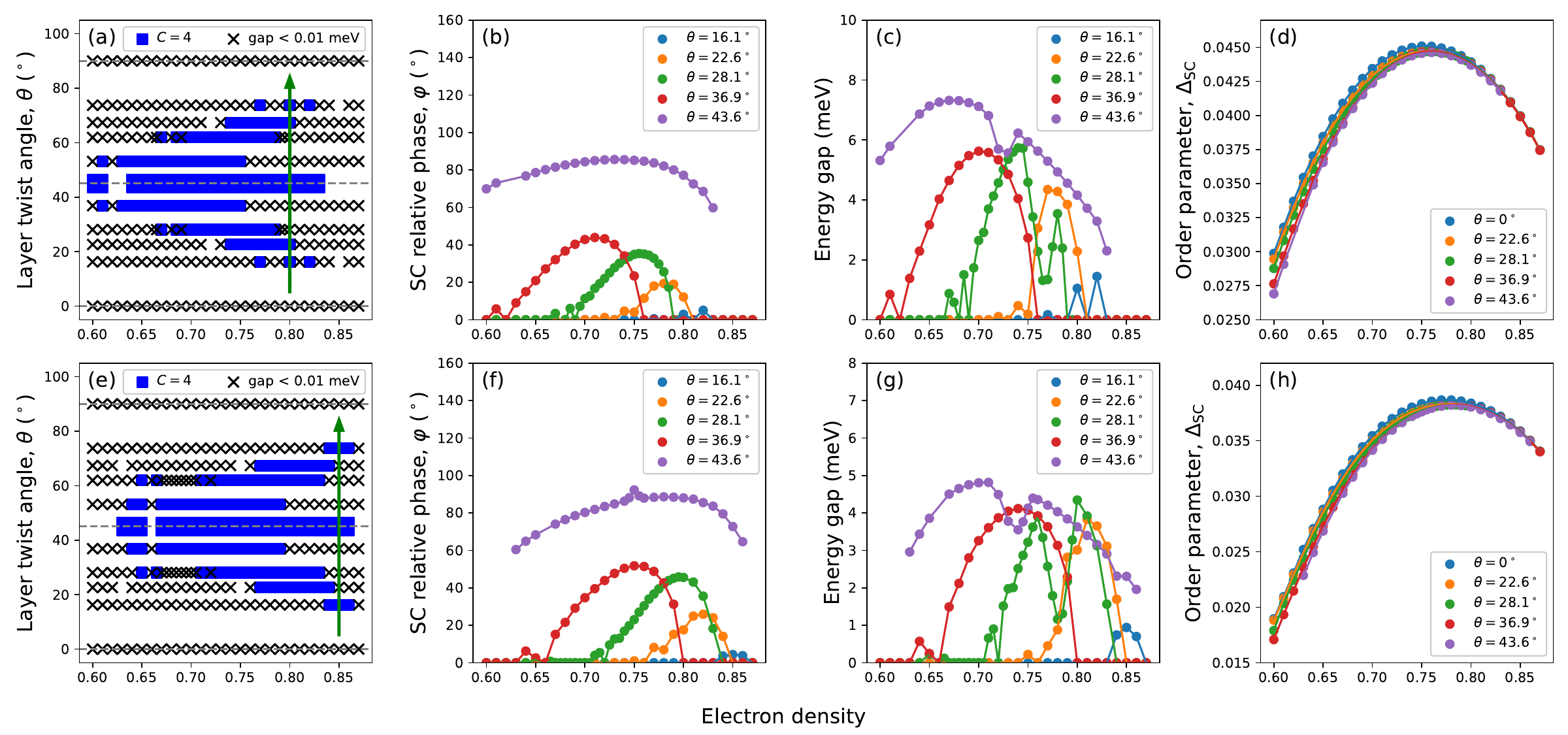}
  \caption{Summary of variational wave function results for twisted square-lattice bilayer $t$-$J$-$U$ model for $J = 0.4 |t|$ (top panels) and $J = 0.3 |t|$ (bottom panels). The remaining model parameters are detailed in the main text. Panels (a) and (e) show the obtained doping vs. twist-angle phase diagrams, with gapless $d$-wave state marked by black crosses (numerically, $\mathrm{gap} < 10^{-2}\,\mathrm{eV}$), and gaped topological SC (Chern number $C=4$). For both values of antiferromagnetic exchange, butterfly-shaped gapped phase is obtained, indicating reentrant TRS-breaking $d+\mathrm{e}^{i\varphi}d$ SC on the underdoped side of the phase diagram. Exemplary constant-density lines, for which reentrant $d+\mathrm{e}^{i\varphi}d$ SC is observed, are marked by green arrows in (a) and (e). The remaining panels show doping-dependence of relative SC phase between layers $\varphi$ [(b) and (f)], energy gap in the quasiparticle spectrum [(c) and (g)], and intraplane SC order parameter $\Delta_\mathrm{SC}$ [(d) and (h)].}
  \label{fig:phase_diag}
\end{figure*}

Before detailed presentation of the obtained phase diagrams, in Fig.~\ref{fig:symmetry_test} we characterize the relevant aspects of superconductivity for  two complementary twist angles, $\theta = 53.13^\circ$ (blue symbols) and $\theta = 36.87^\circ = 90^\circ - 53.13^\circ$ (red symbols), cf. Table~\ref{tab:bilayer_superstructures}. The calculations have been carried out for the $t$-$J$-$U$ model~\eqref{eq:tju-model-layers} with $J = 0.4 |t|$ (left panels) and $J = 0.3 |t|$ (right panels). The remaining model parameters are given in Sec.~\ref{section:model}. From top to bottom, the panels detail  electron-density dependence of the SC order parameter $\Delta_\mathrm{SC}$ [(a) and (d)], equilibrium SC relative phase $\varphi$ [(b) and (e)], and the energy gap defined as the minimum energy for creating a particle-hole excitation [(c) and (f)]. For either selection of $J$, the order parameter [panels (a) and (d)] takes a typical dome-like shape with a maximum close to $20\%$ hole-doping. Remarkably, the SC phase angle $\varphi$ [panels (b) and (e)] also forms a dome along the electron-density axis, yet it encompasses only a fraction of the SC phase diagram. The non-trivial value of $\varphi \neq 0^\circ, 180^\circ$ (modulo $360^\circ$) indicates spontaneous TRS breakdown and emergence of topological SC. The latter is characterized by a non-zero gap in the energy spectrum [cf. panels (c) and (f)] and Chern number $C = 4$. In order to obtain perfectly quantized Chern numbers, we employ efficient Brillouin-zone triangulation scheme \cite{FukuiJPSJ2005} that yields essentially roundoff-error limited results. Such an accuracy is needed to unambiguously characterize weak SC state topology close to the $d+\mathrm{e}^{i\varphi}d$ dome boundary, as evidenced by fractional numerical values of Chern numbers reported within other schemes \cite{SongPhysRevB2022}. On the other hand, for $\varphi = 0^\circ$ or $180^\circ$, the SC state is gapless due to the nodal structure inherited from the layer $d$-wave order parameters. In effect, there are two hidden quantum phase transitions within the SC dome, separating gapless and gapped topological SC (TSC) states. One is located close to optimal doping and the other appears in the overdoped regime; both are marked by dashed vertical lines.

Figure~\ref{fig:symmetry_test} provides also a stringent test of our theoretical framework. The displayed phase diagrams for complementary angles $\theta = 53.13^\circ$ and $\theta = 36.87^\circ = 90^\circ - 53.13^\circ$ have been obtained by independent simulations carried out for different supercells ($n=2, m=1$ and $n=3, m=1$, cf. Table~\ref{tab:bilayer_superstructures}). The collapse of the order parameter and energy gaps indicates physical equivalence of those configurations, as predicted on symmetry grounds. Moreover, the self-consistently obtained equilibrium values of $\varphi$ for those twist angles are compatible with the transformation rules introduced above [cf. panels (b) and (e) of Fig.~\ref{fig:symmetry_test}]; the free energy minima at $\pm \varphi$ for twist angle $\theta$ transforms into $180^\circ \pm \varphi$ at complementary angle $90^\circ - \theta$. Small quantitative differences observed at the boundaries of SC domes originate from finite-size and boundary-condition effects, as explained in Appendix~\ref{appendix:landau_free_energy}.

\subsection{Doping vs. twist angle phase diagrams}

We now generalize the results of Fig.~\ref{fig:symmetry_test} and present the complete doping vs. twist-angle phase diagrams in Fig.~\ref{fig:phase_diag}. Top and bottom panels correspond to antiferromagnetic exchange $J = 0.4 |t|$ and $0.3 |t|$, respectively, with other parameters remaining the same as those listed in Sec.~\ref{section:model}. Black symbols in panels (a) and (e) mark the gapless SC state (numerically, gap below $10^{-2}\,\mathrm{meV}$ is regarded as zero), and blue crosses represent gapped topological $d+\mathrm{e}^{i\varphi}d$ state with Chern number $C=4$. Missing symbols for some combinations of $\theta$ and density indicate that we were unable to reach the absolute precision of $10^{-10}$ for the correlation functions, required to reliably determine order-parameter relative phase, $\varphi$ (cf. the discussion of the computational aspects in Appendix~\ref{appendix:landau_free_energy}. The most distinctive feature of the phase diagrams, displayed in panels (a) and (e), is the non-convex butterfly-shaped region of the TRS-breaking $d+\mathrm{e}^{i\varphi}d$ state for both considered values of $J$. The non-convexity manifests itself as a reentrant behavior of topological SC as a function of bilayer twist angle above optimal doping. Green arrows in panels (a) and (e) mark exemplary constant-density paths inside the phase diagram along which reentrance is observed. Whereas realization of topological SC in the $\theta \sim 45^\circ$ region is obstructed by detailed angular structure of interlayer tunneling \cite{MarkiewiczPhysRevB2005,SongPhysRevB2022}, the $d+\mathrm{e}^{i\varphi}d$ state might emerge for twist-angles in the reentrant-SC range $\theta \sim 20$-$30^\circ$ sufficiently close to half-filling.

\begin{figure}
  \centering
  \includegraphics[width=0.95\linewidth]{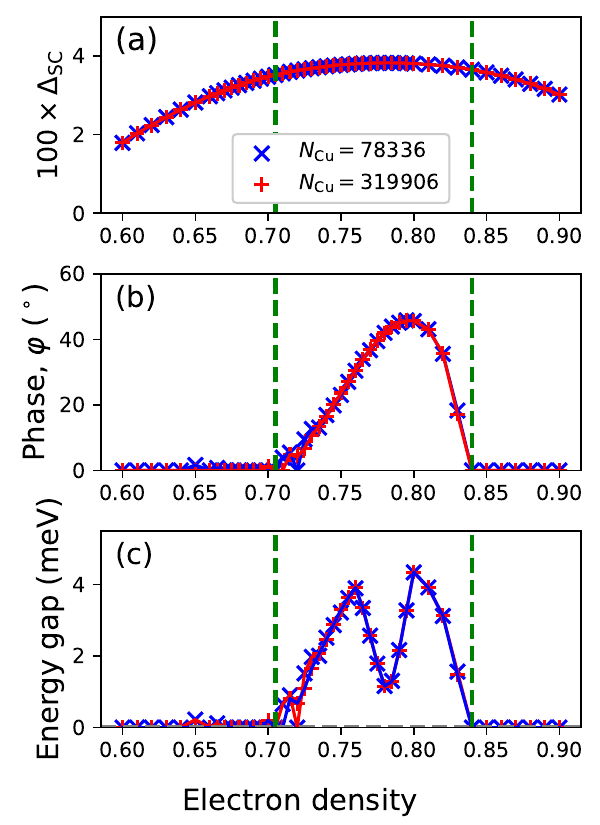}
  \caption{Electronic-density dependence of the principal SC state characteristics, obtained for $J = 0.3 |t|$ and layer twist angle $\theta = 28.07^\circ$. The remaining microscopic model parameters are the same as  those used to generate phase diagrams of Fig.~\ref{fig:phase_diag}. Blue and red curves correspond to the total number of Cu sites $N_\mathrm{Cu} = 78\,336$ and $319\,906$, respectively. Vertical dashed lines mark quantum phase transitions between gapped and gapless states inside the SC dome. All the displayed quantities qualitatively agree for different lattice sizes, i.e. SC order parameter $\Delta_\mathrm{SC}$ (a), relative SC phase $\varphi$ (b), and energy gap (c). In particular, the two-dome structure of energy gap in panel (c) is robust against finite-size scaling, indicating that it is not a numerical artifact.}
  \label{fig:two_sc_domes}
\end{figure}

The remaining part of Fig.~\ref{fig:phase_diag} details the doping-dependence of relevant SC-state characteristics across the phase diagram. Panels (b) and (f) show the equilibrium SC phase angle between layers, $\varphi$. The range of densities, for which nontrivial $\varphi$ is obtained, shrinks and shifts toward half-filling with decreasing $\theta$. Panels (c) and (g) detail the corresponding gaps in the quasiparticle spectrum, varying in the range $0$-$8\,\mathrm{meV}$. The gapped state overlaps with the regime of TRS-breaking $d+\mathrm{e}^{i\varphi}d$ SC. Interestingly though, the relation between energy gaps and $\varphi$ is not straightforward as those two quantities exhibit a qualitatively different doping dependence for specific combination of microscopic parameters. This is clearly seen for $\theta = 28.1^\circ$ and $J = 0.3|t|$ [green symbols in panels (f) and (g)]. Whereas $\varphi$ forms a single dome as function of electron density, the corresponding energy gap splits into two overlapping domes centered at $n \approx 0.75$ and $n \approx 0.80$. To a lesser degree, such a dip in energy gap is also observed at $\theta = 43.6^\circ$. With the use of finite-size scaling, we have verified that the dip in the energy gap for $\theta = 28.1^\circ$ is not a numerical artifact and thus is of physical significance. Indeed, in Fig.~\ref{fig:two_sc_domes} we compare the doping-dependence of SC characteristics for $\theta = 28.1^\circ$ and two lattice sizes, $N_\mathrm{Cu} = 78\,336$ (blue symbols) and $N_\mathrm{Cu} = 319\,906$ (red symbols). All relevant quantities: Order parameter $\Delta_\mathrm{SC}$ (a), phase angle $\varphi$ (b), and energy gaps (c) overlap for different values of $N_\mathrm{Cu}$. In particular, two-dome structure is robust against finite-size scaling, supporting our conclusion. Finally, panels (d) and (h) of the phase diagram (cf. Fig.~\ref{fig:phase_diag}) detail the interlayer $d$-wave order parameter, $\Delta_\mathrm{SC}$, forming a typical superconducting dome. Variation of the twist angle only weakly affects the magnitude of SC correlations within the layers, as those are mostly inherited form the untwisted materials. From the comparison of panels (d) and (h) with panels (a) and (e) of Fig.~\ref{fig:phase_diag}, it becomes apparent that the boundary of reentrant topological SC is positioned close to optimal doping.

\subsection{Landau free energy functional and robustness of topological superconductivity}
\label{subsec:landau_free_energy}

\begin{figure}
  \centering
  \includegraphics[width=0.95\linewidth]{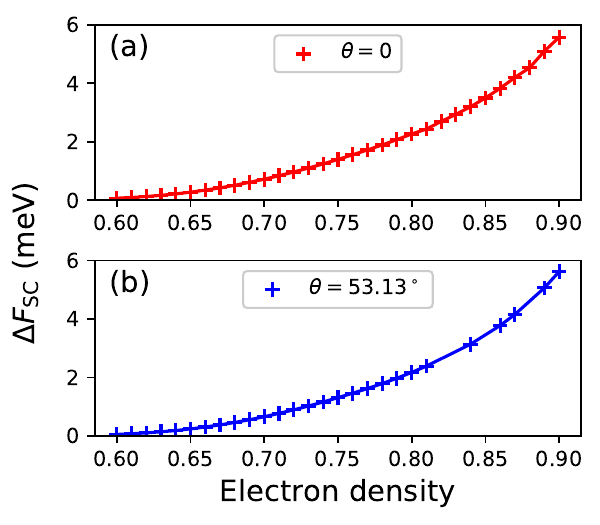}
  \caption{Condensation energy $\Delta F_\mathrm{SC}$ for bilayer $t$-$J$-$U$ model as a function of electron density per copper-site for (a) $\theta = 0^\circ$ and (b) $\theta = 53.13^\circ$. Antiferromagnetic exchange is set to $J = 0.3 |t|$ and the remaining parameters are provided in the main text. The magnitude of $\Delta F_\mathrm{SC}$ varies in single meV range.}
  \label{fig:condensation_energy}
\end{figure}

We now address the hierarchy of energy scales related to SC transition in twisted bilayers, which goes beyond the equilibrium analysis summarized in Figs.~\ref{fig:symmetry_test} and \ref{fig:phase_diag}. For the parent (untwisted) system, the relevant quantity is condensation energy, $\Delta F_\mathrm{SC}$, defined as the difference between the normal- and paired-state free energies per Cu site, $\Delta F_\mathrm{SC} \equiv F_N - F_\mathrm{SC}$. It should be emphasized that there is a degree of arbitrariness to the theoretical treatment of $\Delta F_\mathrm{SC}$ for a high-temperature superconductor as the normal-state free energy, $F_N$, may be identified with either pseudogap phase or correlated Fermi liquid. Here we adopt that latter convention and assume that $F_N$ corresponds to the renormalized Fermi liquid, since pure pseudogap without coexistent SC is not easily described within the present variational scheme. Those aspects, as well as possible relation between pseudogap and SC, have been addressed previously in the context of gossamer SC within a Gutzwiller-type approach \cite{LauglinPhilMag2006}. In effect, thus defined $\Delta F_\mathrm{SC}$ combines contributions attributed to both SC and pseudogap states on the underdoped side of the phase diagram. A monotonic increase of $\Delta F_\mathrm{SC}$ is therefore expected as half-filling is approached, in contrast to the dome-like behavior reported based on thermodynamic measurements \cite{LoramPhysicaC2000,MatsuzakiJPSJ2004}. In Fig.~\ref{fig:condensation_energy} we display calculated $\Delta F_\mathrm{SC}$ as a function of electronic density per Cu atom for two selected layer twist angles, $\theta = 0^\circ$ and $\theta = 53.13^\circ$. In either case, $\Delta F_\mathrm{SC}$ varies within the meV range. Remarkably, $\Delta F_\mathrm{SC} \lesssim 1 \, \mathrm{meV}$ in the overdoped regime, where no pseudogap contribution is present. This is somewhat larger, but within the same order of magnitude as estimates from specific heat measurements for Bi2212 \cite{LoramPhysicaC2000,LevalloisPhysRevX2016}. 

In the case of bilayer, the free energy in the SC state may be regarded as a function of an additional parameter, i.e. relative phase $\varphi$. Variation of the Landau free energy $F(\varphi)$ for $\varphi \in [0, 360^\circ]$ defines the second energy scale $\Delta F_\mathrm{SC}^\prime$. Below, we demonstrate that those two scales obey strict hierarchy $\Delta F_\mathrm{SC}^\prime \ll \Delta F_\mathrm{SC}$. This renders topological SC fragile against small perturbations (e.g., those induced by disorder) and makes realization of homogeneous $d+\mathrm{e}^{i\varphi}d$ state challenging. A methodological remark is in order at this point. Strictly speaking, the thermodynamic system free energy $F$ in not a functional of either $\Delta_\mathrm{SC}$ or $\varphi$. In order to determine $\Delta F_\mathrm{SC}^\prime$, one thus needs to evaluate the Landau free energy functional $F(\varphi)$ as a Legendre transform of the generalized free energy $F(\mathbf{J})$ in the presence of background currents $\mathbf{J}$ coupled linearly to the bilayer $d+\mathrm{e}^{i\varphi}d$ order parameter. This procedure is detailed in Appendix~\ref{appendix:landau_free_energy}.

\begin{figure}
  \centering
  \includegraphics[width=1\linewidth]{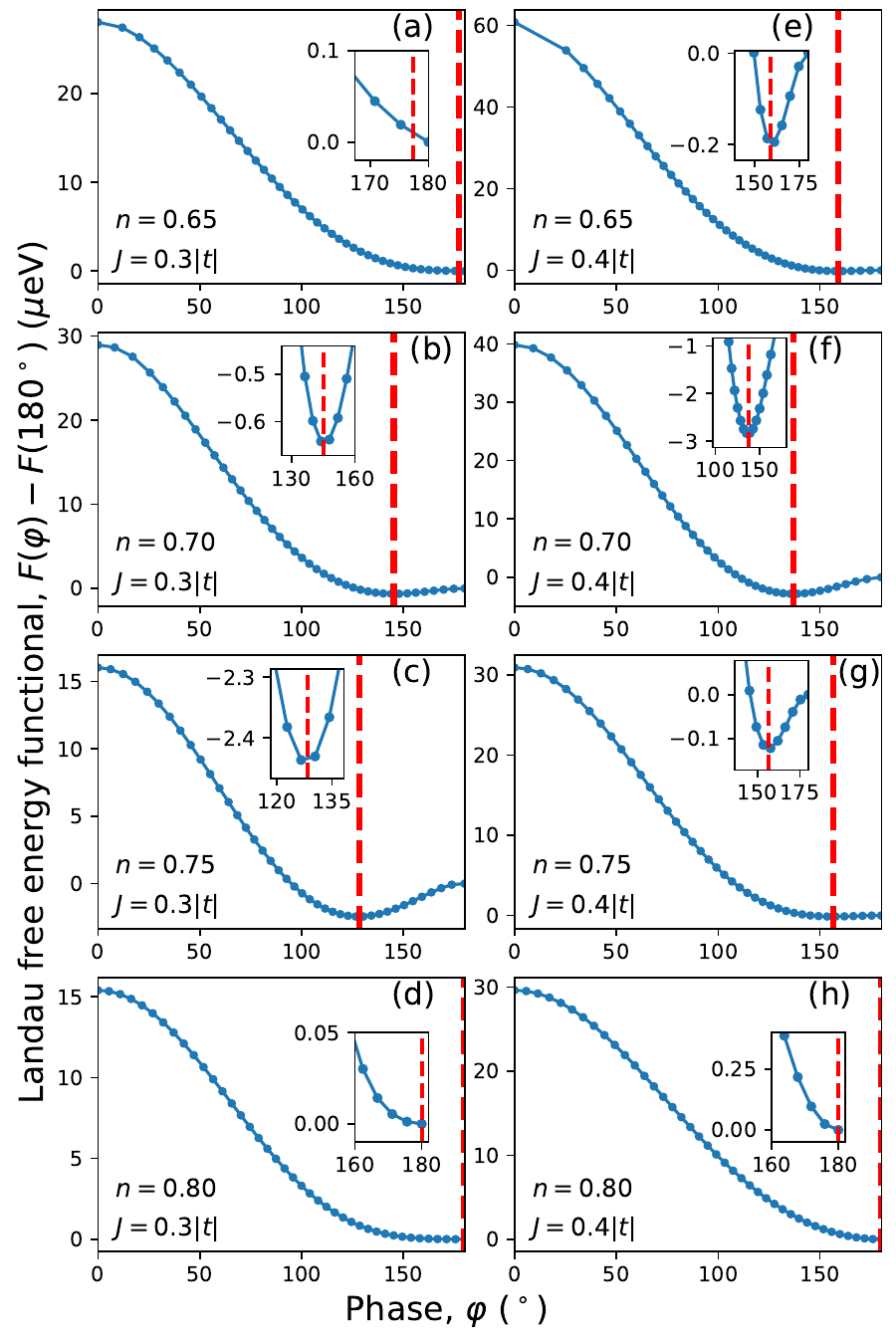}
  \caption{Landau free energy functional, $F(\varphi)$, as a function of relative phase  SC between, $\varphi$. The layer twist angle is set to $\theta = 53.13^\circ$, and values of antiferromagnetic exchange $J$ and electron density $n$ are detailed inside the panels. The remaining parameters are given in the main text. Blue points represent $F(\varphi)$ evaluated for background field magnitude $J_0 = 0.003 |t|$ (cf. Appendix~\ref{appendix:landau_free_energy} for details). Dashed red vertical lines represent the represent exact equilibrium values of phase $\varphi$, obtained by self-consistent SGA calculation. Insets detail $F(\varphi)$ close to the minima, and demonstrate that SGA value of $\varphi$ indeed corresponds to its minimum. Note that the Landau free energy functional varies at the $\mu\mathrm{eV}$ scale and thus requires a careful numerical analysis to achieve reliable results.}
  \label{fig:effective_potential}
\end{figure}

In Fig.~\ref{fig:effective_potential} we plot the Landau free energy functional, $F(\varphi)$, for fixed $\theta = 53.13^\circ$ and various selections of electron density and exchange coupling $J$, covering the phase diagrams displayed in Fig.~\ref{fig:phase_diag}. Left- and right panels correspond to $J = 0.3 |t|$ and $0.4 |t|$, respectively. The unspecified model parameters are the same as those given in Sec.~\ref{section:model}. Points are obtained from SGA calculation, and lines are guides to the eye. Insets inside the panels show the close-up view of $F(\varphi)$ close to minimum, and dashed vertical lines mark the value of $\varphi$ obtained by self-consistent calculation without external currents (and thus representing the true equilibrium solution). 

We note that the minimum of $F(\varphi)$ coincides with the self-consistently obtained value of SC relative phase, $\varphi$. This demonstrates that $F(\varphi)$ represents the appropriate thermodynamic potential and further validates our approach. For $n = 0.65$, $F(\varphi)$ attains minimum close to $\varphi = 180^\circ$, cf. panels (a)-(b). At intermediate densities ($n = 0.70$-$0.75$), Landau free energy is minimized by nontrivial values of $\varphi$ [panels (b)-(e)]. Finally, close to half-filling $n \gtrsim 0.8$, minimum shifts back toward $\varphi = 180^\circ$ [panels (f)-(g)]. This behavior reflects the dome-like structures observed in Fig.~\ref{fig:phase_diag}(b) and (f). Remarkably though, the variation of $F(\varphi)$ in the range $\varphi \in [0, 360^\circ]$ does not exceed $60\,\mu\mathrm{eV}$ which is by two order of magnitude smaller than typical condensation energy for the $t$-$J$-$U$ model in the high-$T_c$ copper-oxide regime. We have thus established $\Delta F_\mathrm{SC}^\prime \ll \Delta F_\mathrm{SC}$.

\section{Summary}
\label{section:summary}

We have studied phase diagram of the $t$-$J$-$U$ model on twisted square lattice, as a function of both hole doping and twist-angle. The symmetry considerations allow us to establish relationship between the solutions for complementary twist angles $\theta$ and $90^\circ - \theta$, which has been utilized to construct a complete superconducting phase diagram using limited number of supercells. The latter mapping is exact in the thermodynamic limit, but controllable finite-size and boundary-condition effects are observed for finite lattices.

The phase diagram comprises both the gapless $d$-wave state and gapped TRS breaking topological $d+\mathrm{e}^{i\varphi}d$ phase. We have found that $d+\mathrm{e}^{i\varphi}d$ state occupies a non-convex butterfly-shaped region in the density vs. twist-angle phase diagram, resulting in reentrance of $d+\mathrm{e}^{i\varphi}d$ pairing as a function of twist angle $\theta$ close to half-filling. One of the footprints of the SC state for non-trivial values of $\varphi$ (i.e.,  $\varphi \neq 0^\circ, 180^\circ$) is the emergence of gap in the quasiparticle energy spectrum, yet the gap magnitude is not related to SC phase angle $\varphi$ in a straightforward manner. In particular, we have identified a multi-dome structure of energy gaps for certain values of layer twist angles. 

The microscopically derived Landau free energy functional exhibits a small variation as a function of the order-parameter relative phase, $\varphi$. In effect, the order parameter is susceptible to small perturbations. We have explicitly demonstrated that finite-size effects and boundary condition effects appear close to the TRS-breaking SC onset, where the free energy landscape is particularly flat. This might rationalize reported difficulties with realization of the homogeneous $d+\mathrm{e}^{i\varphi}d$ state close to twist-angle $\theta = 45^\circ$.

\section*{Acknowledgments}

Two of us (M.F. and J.S.) were partly supported by Grant Opus No. UMO-2021/41/B/ST3/04070 from Narodowe Centrum Nauki (NCN). One of us (B.R.) was entirely supported by the project Opus No. UMO-2018/29/ST3/02646 from NCN. 

\appendix

\section{Statistically-Consistent Gutzwiller approach with on-site SC pairing}
\label{appendix:SGA}

The statistically-consistent variational (Gutzwiller-type) Approximation (SGA) has been formulated and extensively analyzed elsewhere, see Ref.~\citenum{SpalekPhysRep2022,JedrakArXiV2011} and references therein. Here we summarize a specific variant of the latter that incorporates on-site SC pairing. This is needed to properly account for the order-parameter symmetry in lattices with large unit cells, such as those emerging in twisted square-lattice systems.

The basic object within the SGA is variational energy functional

\begin{align}
  \label{eq:variational_energy}
  E_\mathrm{var} = \frac{\langle \Psi_\mathrm{var}|\hat{\mathcal{H}}| \Psi_\mathrm{var}\rangle }{\langle \Psi_\mathrm{var}|\Psi_\mathrm{var} \rangle},
\end{align}

\noindent
where variational wave function is expressed as $|\Psi_\mathrm{var}\rangle \equiv \hat{\mathcal{P}} |\Psi_0\rangle$. Here $|\Psi_0\rangle$ denotes a Slater-determinant state and $\hat{\mathcal{P}}$ is the so-called \emph{correlator} (an operator introducing correlations into the uncorrelated wave function $|\Psi_0\rangle$). We adopt the correlator in the form of a lattice product $\hat{\mathcal{P}} \equiv \prod_i \hat{\mathcal{P}}_i$, where

\begin{align}
  \label{eq:correlator}
  \hat{\mathcal{P}}_i \equiv \lambda_i^0 |0\rangle_i{}_i\langle{0}| + \sum_{\sigma = \uparrow, \downarrow} \lambda_i^\sigma |\sigma\rangle_i{}_i\langle{\sigma}| + \lambda_i^d |d\rangle_i{}_i\langle{d}|.
\end{align}

\noindent
In Eq.~\eqref{eq:correlator}, the projection operators onto the local basis states ($|0\rangle_i$, $|{\uparrow}\rangle_i$, $|{\downarrow}\rangle_i$, and $|d\rangle_i \equiv |{\uparrow\downarrow}\rangle_i$) are multiplied by parameters $\lambda^0_i$, $\lambda^\uparrow_i$, $\lambda^\downarrow_i$, and $\lambda^d_i$. Both the $\lambda$-parameters and uncorrelated state $|\Psi_0\rangle$ are variational objects that need to be determined by minimization of the functional $E_\mathrm{var}$. In particular, $|\Psi_0\rangle$ encodes broken-symmetry states, including unconventional SC. Moreover, additional conditions on the $\lambda$-parameters are needed to make the variational problem tractable, namely

\begin{align}
  \label{eq:constraints}
  \langle \Psi_0| \hat{P}_i^2 |\Psi_0\rangle & = 1, \\
  \langle \Psi_0|\hat{P}_i \hat{n}_{i\uparrow} \hat{P}_i |\Psi_0\rangle & = \langle \Psi_0|\hat{n}_{i\uparrow} |\Psi_0\rangle, \\
  \langle \Psi_0|\hat{P}_i \hat{n}_{i\downarrow} \hat{P}_i |\Psi_0\rangle & = \langle \Psi_0|\hat{n}_{i\downarrow} |\Psi_0\rangle,
\end{align}

\noindent
where $\hat{n}_{i\sigma} \equiv \hat{a}^\dagger_{i\sigma} \hat{a}_{i\sigma}$ is the particle number operator. In effect, the number of $\lambda$-parameters is reduced from four to one per site; without loss of generality, one can select $\lambda^d_{i}$ as the remaining one.

For specified microscopic Hamiltonian, $\hat{\mathcal{H}}$, the functional~\eqref{eq:variational_energy} may be evaluated using Wick's theorem. Formally, variational energy becomes then a functional two-point expectation values of the form $\langle \Psi_0| \hat{a}^\dagger_{i\sigma} \hat{a}_{j\sigma^\prime} |\Psi_0\rangle$, $\langle \Psi_0 | \hat{a}^\dagger_{i\sigma} \hat{a}^\dagger_{j\sigma^\prime} | \Psi_0 \rangle$, $\langle \Psi_0 | \hat{a}_{i\sigma} \hat{a}_{j\sigma^\prime} |\Psi_0 \rangle$, and correlator parameters. For brevity of notation, we denote the set of all two-point correlation function of this form as $\mathbf{P} = (P_1, P_2, \ldots)$ and dub them as \emph{lines}, whereas variational parameters are collectively marked as $\boldsymbol{\lambda} = (\lambda_1, \lambda_2, \ldots)$,  where the indices enumerate all relevant degrees of freedom. Moreover, we introduce analogous notation for the bilinear operators $\hat{\mathbf{P}} \equiv (\hat{P}_1, \hat{P}_2, \ldots)$ composed of the operator products $\hat{a}^\dagger_{i\sigma} \hat{a}_{j\sigma^\prime}$, $\hat{a}^\dagger_{i\sigma} \hat{a}^\dagger_{j\sigma^\prime}$, $\hat{a}_{i\sigma} \hat{a}_{j\sigma^\prime}$ that are related to the corresponding lines as $\langle \Psi_0| \hat{P}_\gamma | \Psi_0 \rangle \equiv P_\gamma$.

In effect, one can write $E_\mathrm{var} \equiv E_\mathrm{var}(\mathbf{P}, \boldsymbol{\lambda})$. Practical methods of evaluating the functional~\eqref{eq:variational_energy} include variational Monte-Carlo or specialized diagrammatic expansions in real space (\textbf{D}iagrammatic \textbf{E}xpansion of the \textbf{G}uztwiller \textbf{W}ave \textbf{F}unction, DE-GWF) \cite{BunemannEPL2012,KaczmarczykNewJPhys2014}. Here, we adopt the latter approach and retain only the leading-order diagrams (those that dominate in the large lattice-coordination-number limit). This results in the so-called \textbf{S}tatistically-\textbf{C}onsistent \textbf{G}utzwiller approximation (SGA) \cite{JedrakPhysRevB2011}. 

We now proceed to formulation of our approach. Variational method amounts to minimization of the functional $E_\mathrm{var} \equiv E_\mathrm{var}(\mathbf{P}, \boldsymbol{\lambda})$ with respect to both $\mathbf{P}$ and $\boldsymbol{\lambda}$, with the additional constraints of fixed electron density and $P_\gamma \equiv \langle\Psi_0| \hat{P}_\gamma | \Psi_0\rangle$. The last condition ensures that the values of lines are compatible with some wave function that belongs to the underlying variational space. We employ a generalization of the plain variational method to finite temperature, based on the free energy functional

\begin{align}
  \label{eq:fre_en_functional}
  \mathcal{F}(\mathbf{P}, \boldsymbol{\lambda}, \boldsymbol{\rho}, \mu) = -\frac{1}{\beta} \ln \mathrm{Tr} \exp(-\beta \hat{\mathcal{H}}_\mathrm{eff}),
\end{align}

\noindent
where

\begin{align}
  \label{eq:effective_hamiltonian}
    \hat{\mathcal{H}}_\mathrm{eff}(\mathbf{P}, \boldsymbol{\lambda}, \boldsymbol{\rho}, \mu) \equiv & E_\mathrm{var}(\mathbf{P}, \boldsymbol{\lambda})+ \sum_\gamma \rho_\gamma \cdot (\hat{P}_\gamma - P_\gamma)  \nonumber \\ & - \mu (\hat{N} - N_e)
  \end{align}

\noindent
is the effective Hamiltonian describing the dynamics of correlated Fermi quasiparticles. In Eq.~\eqref{eq:effective_hamiltonian} $\rho_\gamma$ are Lagrange multipliers ensuring that the values of lines are compatible with thermodynamic expectation values, i.e.,

\begin{align}
  \label{eq:self_consistency_condition}
  P_\gamma = \langle \hat{P}_\gamma \rangle \equiv \frac{\mathrm{Tr} \hat{P}_\gamma \mathrm{e}^{-\beta \hat{\mathcal{H}}_\mathrm{eff}}}{ \mathrm{Tr} \mathrm{e}^{-\beta \hat{\mathcal{H}}_\mathrm{eff}} }.
\end{align}

\noindent
Moreover, the Lagrange multiplier $\mu$ is used to impose that the expectation value of the total particle number operator, $\langle \hat{N}\rangle$, is equal to target electron number $N_e$. 

The system free energy $F$ is determined as a stationary point of the free energy functional over all fields that yield

\begin{align}
  \label{eq:free_energy_minimum_conditions_1}
  \frac{\partial \mathcal{F}}{\partial P_\gamma} =& \frac{\partial E_\mathrm{var}(\mathbf{P}, \boldsymbol{\lambda})}{\partial P_\gamma} - \rho_\gamma \equiv 0, \\
    \label{eq:free_energy_minimum_conditions_2}
  \frac{\partial \mathcal{F}}{\partial \lambda_\gamma} = & \frac{\partial E_\mathrm{var}(\mathbf{P}, \boldsymbol{\lambda})}{\partial \lambda_\gamma}  \equiv 0, \\
 \label{eq:free_energy_minimum_conditions_3}
  \frac{\partial \mathcal{F}}{\partial \rho_\gamma} = & \langle \hat{P}_\gamma \rangle - P_\gamma  \equiv 0,  \\
 \label{eq:free_energy_minimum_conditions_4}
  \frac{\partial \mathcal{F}}{\partial \mu} = & \langle \hat{N} \rangle - N_e  \equiv 0.
\end{align}

\noindent
In effect, we arrive at

\begin{align}
  \label{eq:free_energy}
  F = \mathcal{F}(\tilde{\mathbf{P}}, \tilde{\boldsymbol{\lambda}}, \tilde{\boldsymbol{\rho}}, \tilde{\mu}),
\end{align}

\noindent
where $\tilde{\mathbf{P}}$, $\tilde{\boldsymbol{\lambda}}$, $\tilde{\boldsymbol{\rho}}$, and $\tilde{\mu}$ denote equilibrium (saddle-point) values of the lines, correlator parameters, and Lagrange multipliers. Equations~\eqref{eq:free_energy_minimum_conditions_1}-\eqref{eq:free_energy_minimum_conditions_4} are solved by self-consistent iteration. Due to substantial computational cost involved for large supercells and slow asymptotic convergence for $d+\mathrm{e}^{i\varphi}d$ SC state in the strong-coupling limit, we have modified the usual self-consistent loop by employing Anderson acceleration scheme \cite{HendersonJComGraphStat2019}.

\begin{figure}
  \centering
  \includegraphics[width=1\linewidth]{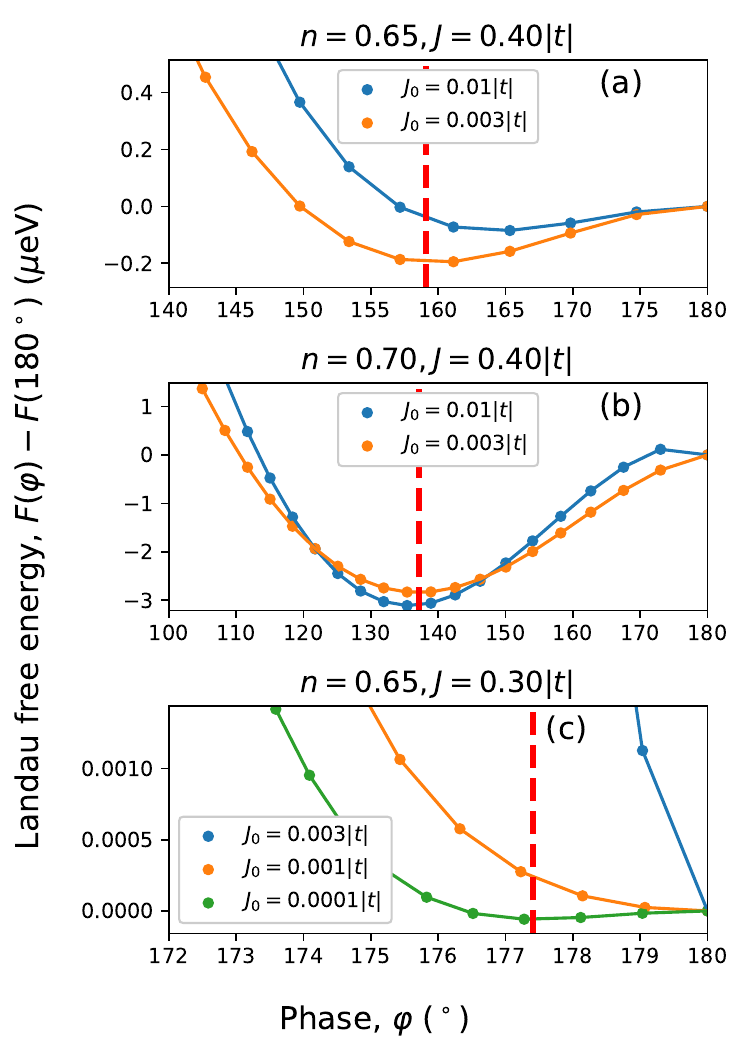}
  \caption{Dependence of the Landau free energy functional on the relative superconducting phase between layers, $\varphi$, for various selections of background current magnitude $J_0$ (cf. Appendix~\ref{appendix:landau_free_energy} for definition). The electronic density and exchange coupling are set to $n = 0.65$, $J = 0.4 |t|$ [panel (a)], $n = 0.70$, $J = 0.4 |t|$ [panel (b)], and $n = 0.65$, $J = 0.3 |t|$ [panel (c)]. The remaining parameters are the same as those used in the main text. Vertical dashed lines mark the self-consistently calculated SGA values for $J_0 \equiv 0$, representing true equilibrium solution.  In the regime of large $J$ and robust $d+\mathrm{e}^{i\varphi}d$ SC, the equilibrium value is reached already for $J_0 = 0.003 |t|$. However, for the smaller values of $J$ and close to topological phase boundary [panel (c)] effective potential becomes practically flat (cf. the energy scale) and then much smaller values $J_0$ are required.}
  \label{fig:effective_potential_details}
\end{figure}

Equivalence of the thermal variational problem based on Eq.~\eqref{eq:fre_en_functional} and plain zero-temperature calculation may be  established by  recasting Eq.~\eqref{eq:free_energy} in a different form

\begin{align}
  \label{eq:free_energy_and_entropy}
  F = E_\mathrm{var}(\tilde{\mathbf{P}}, \tilde{\boldsymbol{\lambda}}) - T S,
\end{align}

\noindent
where the entropy reads

\begin{align}
  \label{eq:entropy}
  S = - k_B \sum_\alpha \Big[\tilde{n}_\alpha \ln \tilde{n}_\alpha +  (1- \tilde{n}_\alpha) \ln (1 - \tilde{n}_\alpha) \Big].
\end{align}

\noindent
In Eq.~\eqref{eq:entropy} the summation index $\alpha$ enumerates eigenstates of the effective Hamiltonian, $\hat{\mathcal{H}}_\mathrm{eff}$, and $\tilde{n}_\alpha$ denotes their equilibrium occupation numbers. By taking $T \rightarrow 0$ the problem is thus reduced to optimization of the plain energy functional, and the thermal expectation values reduce to ground-state averages.

\begin{figure*}
  \centering
  \includegraphics[width=1\linewidth]{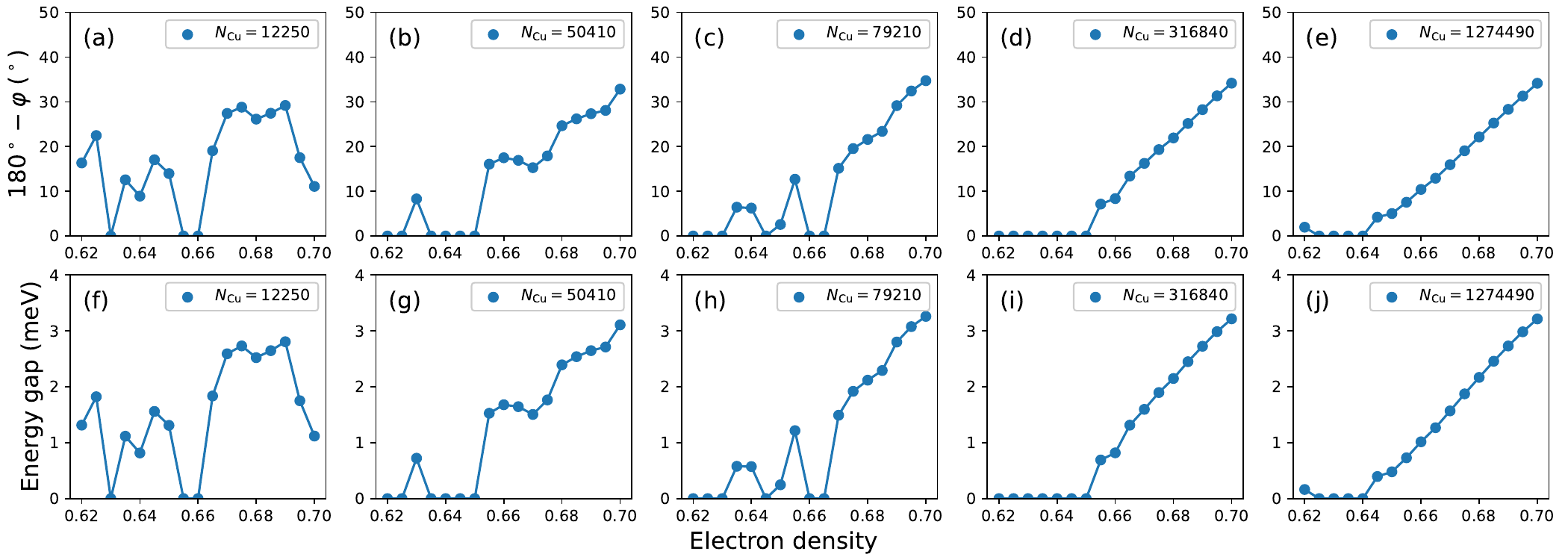}
  \caption{Illustration of the finite-size effects on topological superconducting state in twisted cuprates for twist angle $\theta = 53.13^\circ$ and $J = 0.3 |t|$. The remaining model parameters are listed in Sec.~\ref{section:model} of the main text. Panels (a)-(e) represent doping-dependence of the relative order parameter phase $\varphi$ for various total system sizes spanning two orders of magnitude (employed number of copper sited, $N_\mathrm{Cu}$, is detailed inside the panels). Panels (f)-(j) show the corresponding values of energy gap. Starting from $N_\mathrm{Cu} \sim 80\,000$ superconductivity with gaps $> 1\,\mathrm{meV}$ stabilizes and reflects thermodynamic-limit behavior, but small finite-size effects are still visible in the sub-meV energy range.}
  \label{fig:finite_size_scaling}
\end{figure*}

\section{Landau free energy functional and finite-size effects}
\label{appendix:landau_free_energy}

The free energy, $F$, is a function of temperature and microscopic model parameters (including hopping integrals, Coulomb repulsion, and antiferromagnetic exchange), but does not depend on the superconducting order parameter, cf. Eq.~\eqref{eq:free_energy}. In order to analyze the evolution of the energetic landscape as a function of twist angle, one thus needs to carry out Legendre transform of $F$ and evaluate \emph{effective potential} with respect to auxiliary currents coupled to $d+\mathrm{e}^{i\varphi}d$ order parameter. We follow the condensed-matter convention, and hereafter refer to it as Landau free energy functional.

First, we define the extended Hamiltonian

\begin{align}
  \label{eq:extended_hamiltonian}
  \hat{\mathcal{H}}^\prime(\mathbf{J}) \equiv \hat{\mathcal{H}} - \mathbf{J}^{\dagger} \hat{\boldsymbol{\Delta}} - \hat{\boldsymbol{\Delta}}^\dagger \mathbf{J},
\end{align}
where $\mathbf{J} \equiv (J^{(A)}, J^{(B)})^T$ is a two-component complex external pairing field and $\hat{\boldsymbol{\Delta}} \equiv (\hat{\Delta}^{(A)}, \hat{\Delta}^{(B)})^T$ is the $d+\mathrm{e}^{i\varphi}d$ SC operator for twisted bilayer [cf. Eqs.~\eqref{eq:delta_a}-\eqref{eq:delta_b}]. The superscript $T$ indicates transposition. For compactness, we write $\hat{\mathcal{H}}(\mathbf{J})$ rather than $\hat{\mathcal{H}}(\mathbf{J}, \mathbf{J}^\dagger)$, keeping in mind that $\mathbf{J}$ is complex and $\hat{\mathcal{H}}$ depends both on $\mathbf{J}$ and $\mathbf{J}^\dagger$ so that it remains manifestly Hermitian, cf. Eq.~\eqref{eq:extended_hamiltonian}. The same shorthand notation is employed for all functionals introduced below.

One can now repeat the procedure described in Appendix~\ref{appendix:SGA} for the extended Hamiltonian $\hat{\mathcal{H}}^\prime(\mathbf{J})$, and evaluate the corresponding free energy $F^\prime(\mathbf{J})$ as a function of pairing field, $\mathbf{J}$. The expectation value of the order parameter depends on $\mathbf{J}$ and may be expressed as

\begin{align}
  \label{eq:delta_vs_j_1}
  \boldsymbol{\Delta} \equiv -\frac{\partial F^\prime(\mathbf{J})}{\partial \mathbf{J}^\dagger} \\
  \label{eq:delta_vs_j_2}
  \boldsymbol{\Delta}^\dagger \equiv -\frac{\partial F^\prime(\mathbf{J})}{\partial \mathbf{J}} 
\end{align}

\noindent
The Landau free energy is defined as

\begin{align}
  \label{eq:effective_potential}
  F(\boldsymbol{\Delta}) \equiv F^\prime[\mathbf{J}(\mathbf{\Delta})] + \mathbf{J}^\dagger({\boldsymbol{\Delta}}) \boldsymbol{\Delta} +  \boldsymbol{\Delta}^\dagger \mathbf{J}({\boldsymbol{\Delta}}),
\end{align}

\noindent
where the dependence of background currents on the order parameter, $\mathbf{J}({\boldsymbol{\Delta}})$ and $\mathbf{J}^\dagger({\boldsymbol{\Delta}})$, is obtained by inverting Eqs.~\eqref{eq:delta_vs_j_1} and \eqref{eq:delta_vs_j_2}.

\begin{figure}
  \centering
  \includegraphics[width=1\linewidth]{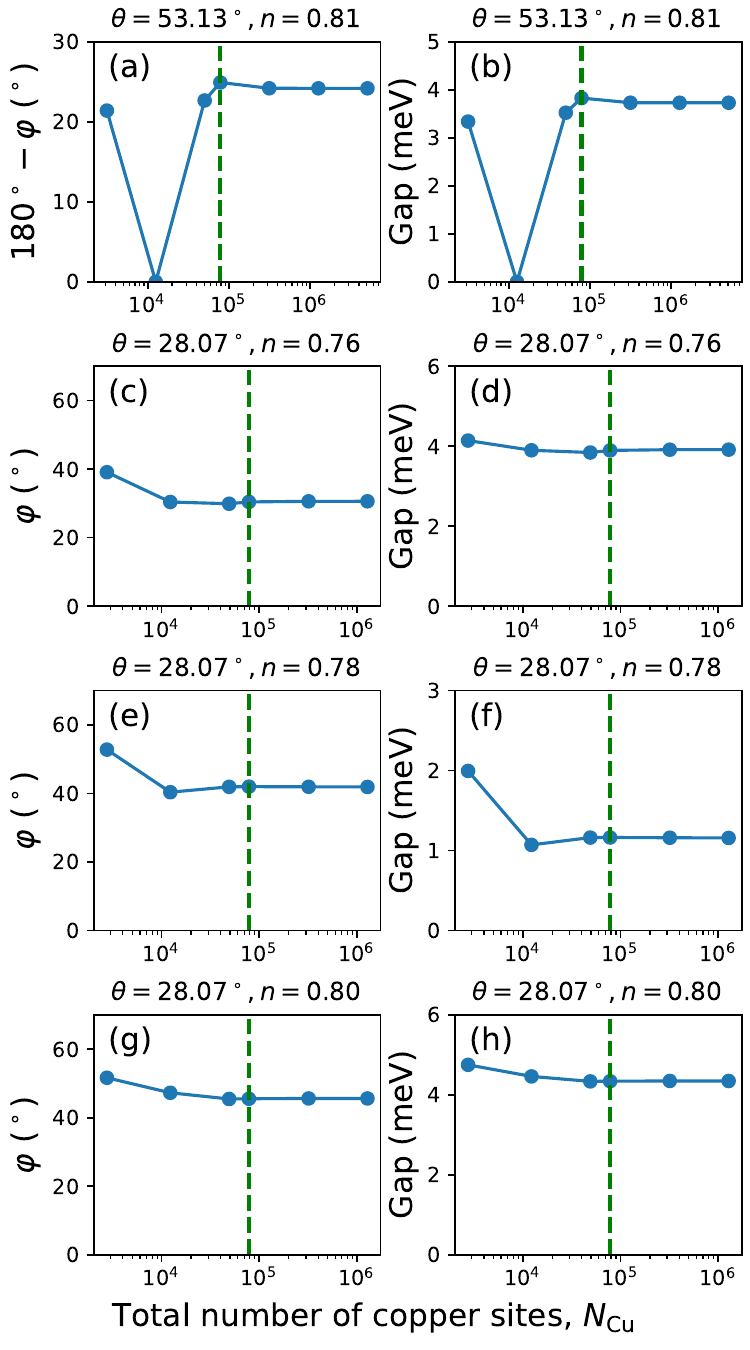}
  \caption{Finite-size scaling of topological superconducting domes for antiferromagnetic exchange $J = 0.3 |t|$; the remaining model parameters are the same as those listed in Sec.~\ref{section:model} of the main text. Panels (a) and (b) show the dependence of the order-parameter phase angle $\varphi$ and energy on the total number of copper atoms, $N_\mathrm{Cu}$, for the twist angle $\theta = 53.13^\circ$ and electron concentration $n = 0.81$. Dashed vertical lines mark the system size ($\approx 80\,000$ Cu sites) used to generate the phase diagrams presented in the main text. The remaining panels (c)-(h) show an analogous scaling carried out for twist angle $\theta = 28.07^\circ$ and doping levels $n = 0.76$, $0.78$, and $0.80$ (encompassing the uncommon double superconducting dome seen in Fig.~\ref{fig:phase_diag}(g) for this value of $\theta$). The finite-size scaling proves that the results displayed in Fig.~\ref{fig:phase_diag} accurately represent the thermodynamic limit situation. The twin-dome behavior is thus not an artifact, but reflects the physics of bulk twisted bilayer system (cf. Fig.~\ref{fig:phase_diag}(g) and panels (d), (f), and (g) of the present figure).}
  \label{fig:finite_size_scaling_dome}
\end{figure}

As is apparent form Eq.~\eqref{eq:effective_potential}, the  Landau free energy is a functional a complex two-component field, $\boldsymbol{\Delta}$, which yields four real degrees of freedom. Since evaluation of $F(\boldsymbol{\Delta})$ is a numerically expensive task, we now propose a way to reduce the number of free parameters  from four to one. First, it should be noted that $F(\boldsymbol{\Delta})$ is invariant with respect to a global gauge transformation $\boldsymbol{\Delta} \rightarrow \exp(i\phi) \boldsymbol{\Delta}$. In turn, the global phase angle $\phi$ might be outright eliminated so that only three nontrivial degrees of freedom in the order-parameter $\boldsymbol{\Delta}$ remain. Without loss of generality, one can select the layer SC amplitudes $|\Delta^{(A)}|$ and $|\Delta^{(B)}|$, and relative phase, $\varphi = \arctan (\Delta^{(B)}/\Delta^{(A)})$. The final simplification is based on the circumstance that there the energy scales related to the SC transition and variation of relative phase $\varphi$ are well separated and may be analyzed independently (cf. discussion in Sec.~\ref{subsec:landau_free_energy}). We thus fix the pairing amplitudes $|\Delta^{(A)}|$ and $|\Delta^{(B)}|$ by imposing two constraints on the amplitudes of background currents in Eqs.~\eqref{eq:delta_vs_j_1}-\eqref{eq:delta_vs_j_2}, namely $|J^{(A)}| = J_0$ and $|J^{(B)}| = J_0$ with $J_0$ being a small positive number. Strictly speaking, the SC amplitudes attain their equilibrium values for $J_0 \rightarrow 0$ or, equivalently, $\mathbf{J} = \mathbf{0}$. Indeed, by combining Eqs.~\eqref{eq:free_energy_minimum_conditions_1}-\eqref{eq:free_energy_minimum_conditions_2} and~\eqref{eq:effective_potential}, one can verify that

\begin{align}
  \label{eq:derivative_landau_free_energy}
  \frac{\partial F(\boldsymbol{\Delta})}{\partial \boldsymbol{\Delta}^\dagger}  = \mathbf{J},
\end{align}

\noindent
so the necessary condition for the Landau free energy minimum $\frac{\partial F(\boldsymbol{\Delta})}{\partial \boldsymbol{\Delta}^\dagger} = \mathbf{0}$ is equivalent to vanishing of the background currents, $\mathbf{J}$. However, keeping a small finite $J_0$ improves performance and ensures stability of solutions in broad range of $\varphi$, so we typically retain non-zero value of $J_0$ in our calculations while evaluating $F(\boldsymbol{\Delta})$. Based on numerical experiments, we have found that $J_0 = 0.003 |t|$ is sufficiently small not to alter the equilibrium value of $\varphi$ for most microscopic parameter configurations considered, and allows to efficiently map the free energy landscape as a function of SC relative phase angle.

In Fig.~\ref{fig:effective_potential_details} we analyze the scaling of $F(\varphi)$ as a function of $J_0$ for selected values of electron density and antiferromagnetic exchange coupling, $J$, detailed above the panels. All other parameters are listed in Sec.~\ref{section:model} of the main text. The vertical dashed lines show the values of relative SC phase $\varphi$, obtained for self-consistent iteration with $J_0 = 0$ and thus reflecting the equilibrium state. As is apparent from Fig.~\ref{fig:effective_potential_details}(a) for $J_0 = 0.01 |t|$, the functional $F(\varphi)$ attains its minimum at non-trivial angle $\varphi \approx 165^\circ$ (blue curve) that differs from true equilibrium value $\varphi \approx 159^\circ$ (vertical red line). By reducing $J_0$ to $0.003 |t|$, the minimum of the Landau free energy shifts to its equilibrium position. This shows that $J_0 \lesssim 0.003 |t|$ already provides a reliable representation of the $J_0 \rightarrow 0$ limit situation. Parenthetically, the scaling presented in Fig.~\ref{fig:effective_potential_details}(a) corresponds to the lower boundary the topological $d+\mathrm{e}^{i\varphi}d$ state for $\theta = 53.13^\circ$ and $J = 0.4 |t|$ [cf. the dome for complementary angle $36.87^\circ$ in the phase diagram of Fig.~\ref{fig:phase_diag}(b) and (c)]. In Fig.~\ref{fig:effective_potential_details}(b), we show an analogous scaling deep inside the dome formed by topological SC state. In this case, we conclude that, already for $J_0 = 0.01 |t|$, the exact position of the Landau free energy minimum (i.e. that obtained by self-consistent iteration for $J_0 = 0$) is accurately determined, and the Landau free energy minimum is deeper than the corresponding one close to the onset of gapped state [panel (a)]. We also note that the exact value of the relative SC phase $\varphi$ is highly sensitive to even small change of $J_0$ on the boundary of the topological state and becomes fairly robust with the increase of energy gap. In panel (c), we demonstrate an extreme case of  such a fragile topological state for fine tuned set of parameters close to the $d+\mathrm{e}^{i\varphi}d$ dome boundary ($n = 0.65 |t|$ and $J = 0.3 |t|$). Here $J_0$ as low as $0.0001 |t|$ is needed to match the self-consistently obtained relative phase. The situation presented in Fig.~\ref{fig:effective_potential_details}(c) is not common and occurs only in regime of very weak topological SC. The free energy landscape is then particularly flat (variation of $F(\varphi)$ within the range of $\mathrm{neV}$), contrary to typical cases displayed in Fig.~\ref{fig:effective_potential_details}(a)-(b), where $F(\varphi)$ varies at the $\mu \mathrm{eV}$ scale.

The above consideration allow us to draw a few general conclusions. First, due to flatness of the free energy landscape, topological superconductivity is expected to by highly sensitive to boundary conditions and finite-size effects at the topological state boundary, even for relatively large lattices. This is clearly seen in Fig.~\ref{fig:phase_diag}(b)-(c) and (f)-(g), where the equilibrium phase and energy gap becomes noisy at the SC dome corners. However, deep within the $d+\mathrm{e}^{i\varphi}d$ state, no noise is observed. Below we carry out a systematic finite-size scaling to further explore those effects. Second, reliable evaluation of equilibrium relative phase $\varphi$ requires high numerical accuracy. In the present paper, we have set the target absolute precision for the dimensionless two-point correlation functions to $10^{-10}$ to achieve this goal. With the number of integral equations to be solved self-consistently exceeding $10^3$ for largest considered supercells ($\theta = 43.6^\circ$), this makes the problem computationally challenging. In particular, for certain densities and $\theta = 43.6^\circ$ in the phase diagrams of Fig.~\ref{fig:phase_diag}(a) and (e), we were unable to obtain the SGA solution with the target accuracy of $10^{-10}$ (missing points). This happens in particular at small doping, where the correlation effects increase as the metal to insulator transition is approached.

Now we address in detail the finite size effects by focusing on layer twist angle $\theta = 53.13^\circ$ and $J = 0.3 |t|$. The remaining model parameters are presented in Sec.~\ref{section:model} of the main text. Figure~\ref{fig:finite_size_scaling} shows the electron-density-dependence of the relative SC phase $\varphi$ [panels (a)-(e)] and energy gap in the quasi particle spectrum [panels (f)-(j)], for various lattice sizes increasing from left to right. The exact number of copper sites in the system, $N_\mathrm{Cu}$, is listed inside the panels and spans two orders of magnitude (from $12\,250$ to $1\,274\,490$). The electron-densities cover the lower-end of the topological $d+\mathrm{e}^{i\varphi}d$ state, where finite-size effects are expected to be particularly relevant, cf. the discussion in Appendix.~\ref{appendix:landau_free_energy}. As follows from Fig.~\ref{fig:finite_size_scaling}, both $\varphi$ and energy gap fluctuates substantially as a function of electron concentration for the smallest lattice sizes considered. The underlying reason is flatness of the Landau free energy functional, so that even fairly weak finite-size effects may affect the equilibrium value of relative SC phase. Above $N_\mathrm{Cu} \sim 80\,000$, both $\varphi$ and gaps start to stabilize, also in the sub-meV range. We thus consider $N_\mathrm{Cu}$ as a threshold value that may be regarded as a representation of the thermodynamic-limit situation. The results reported in the main text have been obtained with lattices containing $N_\mathrm{Cu} \approx 80\,000$ sites.

To further investigate saturation of the topological SC characteristics with increasing lattice size, in Fig.~\ref{fig:finite_size_scaling_dome} we carry out finite size-scaling for $J = 0.3 |t|$, and twist angles $\theta = 53.13^\circ$ and $\theta = 28.07^\circ$, across the $d+\mathrm{e}^{i\varphi}d$ SC dome. The relevant parameters are listed above the panels, and those unspecified are given in Sec.~\ref{section:model} of the main text. Dashed vertical lines mark the lattice sizes $N_\mathrm{Cu} \approx 80\,000$, used to generate the phase diagram of Fig.~\ref{fig:phase_diag}. The finite-size scaled quantities exhibit saturation above $N_\mathrm{Cu} \approx 80\,000$. We also point out that the results displayed in Fig.~\ref{fig:finite_size_scaling_dome}(d), (f), and (h) provide robust evidence for the existence of the two-dome structure [cf. green curve in Fig.~\ref{fig:phase_diag}(g)]. Indeed, the saturated value of the energy gap initially decreases from $\approx 4 \, \mathrm{meV}$ to $\approx 1 \, \mathrm{meV}$ as electron density changes from $n = 0.76$ to $n = 0.78$, and then increases again to $\approx 4 \, \mathrm{meV}$ at  $n = 0.80$. 

\newpage

\input{article.bbl}
\end{document}

%% file: article.bbl
%